\documentclass[twocolumn,secnumarabic,amssymb, nobibnotes, aps, prc, superscriptaddress,nobalancelastpage,nofootinbib]{revtex4-1}

\usepackage{bm}
\usepackage{amsmath} \usepackage{braket} \usepackage{epsfig}
\usepackage{tensor}
\usepackage[version=4]{mhchem}
\usepackage{titlesec}
\usepackage{ifthen}
\usepackage{sidecap}
\usepackage{listings}
\usepackage[para,online,flushleft]{threeparttablex}
\usepackage{mathtools}
\usepackage{siunitx}
\usepackage[normalem]{ulem}
\usepackage{natbib}

\usepackage{soul}

\usepackage{xr-hyper}
\usepackage{hyperref}
\hypersetup{breaklinks=true,colorlinks=true,linkcolor=blue,citecolor=blue,filecolor=magenta,urlcolor=cyan}

\usepackage[all]{hypcap}
\usepackage{xcolor}

\newcommand{\be}{\begin{equation}} 
\newcommand{\ee}{\end{equation}}
\newcommand{\mm}{\mathrm}

\begin{document}

\title{Synthesis of super-heavy elements in the outer crust of a magnetar} 

\author{D. Basilico}
\affiliation{Dipartimento di Fisica ``Aldo Pontremoli'', Universit\`a degli Studi di Milano, 20133 Milano, Italy}
\affiliation{INFN, Sezione di Milano, 20133 Milano, Italy}

\author{G. Col\`o}
\affiliation{Dipartimento di Fisica ``Aldo Pontremoli'', Universit\`a degli Studi di Milano, 20133 Milano, Italy}
\affiliation{INFN, Sezione di Milano, 20133 Milano, Italy}

\author{Xavier Roca-Maza}
\affiliation{Dipartimento di Fisica ``Aldo Pontremoli'', Universit\`a degli Studi di Milano, 20133 Milano, Italy}
\affiliation{INFN, Sezione di Milano, 20133 Milano, Italy}
\affiliation{Departament de F\'isica Qu\`antica i Astrof\'isica, Mart\'i i Franqu\'es, 1, 08028 Barcelona, Spain}
\affiliation{Institut de Ci\`encies del Cosmos, Universitat de Barcelona, Mart\'i i Franqu\'es, 1, 08028 Barcelona, Spain}

\date{\today}
\begin{abstract}
  A theoretical understanding of a possible mechanism for synthesizing super-heavy elements in the outer crust of magnetars is presented. We demonstrate that such a mechanism can be present whenever the baryon density in the outer crust of a neutron star reaches values around $10^{-2}$ fm$^{-3}$. This scenario could be realized in magnetars with hypothetical large magnetic fields, $B\gtrsim 10^{18}$ G. Under such conditions, the Coulomb lattice, formed by nuclei, enables a mechanism that synthesizes super-heavy elements.  
\end{abstract}

\maketitle

\section{Introduction}
Neutron stars are one of the most fascinating objects in the Universe \cite{Haensel:2007yy}.  Among their many unique features, they host extremely strong magnetic fields whose intensity can be in the range of $\approx 10^{11}-10^{14}$ G. There is no understanding of either the origin or the structure of these fields, although the simple argument of the magnetic flux conservation during the collapse of a progenitor main sequence star suggests values of the order of $\approx 10^{12}$ G. From observations of the last decade, values up to $2.4\times 10^{15}$ G have been deduced \cite{Seiradakis2004,Ng:2011,Mereghetti:2008,Olausen_2014,Tiengo2013}. However, even stronger values cannot be ruled out, and possible magnetic field intensities up to $\approx 10^{18}$ G have been suggested, e.g., in Refs. \cite{Stella:2005yz,Potekhin:1997,Potekhin:1999}. The physics of stars characterised by a strong magnetic field such as the magnetars, which are a subset of neutron stars, is an active field of research, and understanding the nature and implications of this huge magnetization is a challenge for astronomy and astrophysics~\cite{Kaspi:2017fwg}.

The outer crust of a neutron star is composed of nuclei arranged in a Coulomb lattice surrounded by a relativistic electron gas~\cite{Shapiro,Haensel:2007yy}. While the lattice is unaltered by the presence of magnetic fields~\cite{vleck1932}, the electron energy levels are quantized according to the famous Landau-Rabi expressions~\cite{Rabi1928,1930ZPhy...64..629L}. This quantization crucially affects the electronic contribution to the density and pressure in the outer crust if extremely large magnetic fields are present. In such conditions, the nuclear structure may be also affected. Indeed, external magnetic fields stabilize nuclei, producing an extra binding that does not exceed 10\%-15\% when the more extreme magnetic fields, $B\approx 10^{18}$ G, are taken into account~\cite{Pena:2011, Basilico:2015ypa, Wei:2024huk}. 

The magnetic field in a magnetar cannot be arbitrarily high: by using the virial theorem, one can qualitatively estimate the upper limit of the neutron star magnetic fields, which is, indeed, of the order of $10^{18} \mathrm G$~\cite{Lai:1991}. It is worth noting that, in the present work, we do not make any assumption on the geometry of the magnetic field throughout the whole outer crust. Indeed, the magnetic field spatial configuration in a magnetar is still an open and interesting problem. Several works~\cite{Tayler1973,TaylerMarkey1973,Wright1973,BraithwaiteSpruit2006,Braithwaite2009,Becerra2022} mainly exclude some field configurations (e.g. pure poloidal or pure toroidal) but do not offer strong conclusions on a unique field geometry at the outer crust scale. However, this open question regarding the spatial configuration of the magnetic field is not, in itself, a strong limitation of this study. We just assume that the magnetic field is locally constant. Our goal is to highlight a mechanism for the synthesis of superheavy nuclei that may take place provided that this local magnetic field is strong enough.

Which nuclei can be precisely synthesized, how does the magnetic field affect this kind of ``nucleosynthesis'', and what can be learnt therefrom? These are the key questions that have already been the subject of a few investigations. In particular, the works that have been published since the last decade and devoted to the effect of the magnetic field on the composition of the outer crust include Refs. \cite{Pena:2011,Chamel2012,ChamelN.2015RoLq,Basilico:2015ypa,2023PhRvD.107d3022P,Chamel2020b,Wei:2024huk}. Recently, the authors of Ref.~\cite{sekizawa2023} have explored extremely large magnetic fields $B\gtrsim 10^{18}$ G and suggested that, in the innermost layers of the outer crust, with average baryon densities of about $10^{-2}$ fm$^{-3}$ (cf. Fig.~3 of Ref.~\cite{sekizawa2023}), super-heavy elements (SHEs) show up. The properties of the outer crust are known to depend on the magnetic field. Indeed, in Ref.~\cite{Chamel2012}, it is shown that both the surface density as well as the neutron drip density -- defining the separation with the inner crust -- could undergo large changes if extreme magnetic fields are present. For example, for $B=0$, the density range spans seven orders of magnitude: from $\approx 10^{-11}$ fm$^{-3}$, i.e. the minimum value for which the complete ionization of atoms in the surface layers of a neutron star takes place, to $\approx 10^{-4}$ fm$^{-3}$ i.e. the neutron drip \cite{Shapiro, Haensel:2007yy}. For extremely large magnetic fields such as $B\sim 10^{18}$ G, this range is shrunk and shifted to larger densities [cf. Eqs.~(27) and (51) as well as Fig.~2 in Ref.~\cite{Chamel2012}, Fig.~6 in Ref.~\cite{Basilico:2015ypa}, Fig.~\ref{fig:EOS_UNEDF1} and Fig.~\ref{fig:EOS_DDPC1} in Sec.~\ref{sec:res}] reaching neutron drip densities of about $10^{-2}$ fm$^{-3}$. 
Consequently, we can say that the presence of densities that may trigger the appearance of SHEs is intimately related to the existence of a strong magnetic field. 

The findings of Ref. \cite{sekizawa2023}, although very interesting, are not explained in simple terms. Our motivation in the present work is twofold. Not only do we want to assess if the results of \cite{sekizawa2023} are confirmed if one uses different state-of-the-art nuclear mass models, but we also aim to understand the reason for the synthesis of SHE in simple and yet robust terms, in a qualitative and as model-independent as possible manner.

With that in mind, in Sec.~\ref{sec:theo}, we will briefly remind the theoretical general
framework, but we shall also present a model for the qualitative understanding of the properties of the matter in the outer crust, valid for average baryon densities $n\sim 10^{-2}$ fm$^{-3}$ and extremely large magnetic fields $B\gtrsim 10^{18}$ G. This will allow us to understand the underlying physical mechanism that may produce super-heavy elements in simple yet reliable terms. In Sec.~\ref{sec:res} we will confirm our understanding by 
discussing numerical results obtained with two state-of-the-art nuclear models. Our conclusions will be laid in Sec.~\ref{sec:con}.

\section{Theoretical model}
\label{sec:theo}
The outer crust is assumed to be composed of nuclei in their ground state at zero temperature, arranged in a Coulomb lattice, and of a relativistic electron gas~\cite{Shapiro,Haensel:2007yy,Pena:2011,Chamel2012,ChamelN.2015RoLq,Basilico:2015ypa,2023PhRvD.107d3022P,Chamel2020b,Wei:2024huk}. We consider this system to be embedded in a magnetic field that extends throughout the outer crust; for our purposes, from now on we will simply consider the magnetic field strength to be locally constant, without any assumptions about its geometry in the outer crust itself. The equilibrium in the outer crust is established by demanding that the temperature ($T$), the pressure ($P$), and the chemical potential ($\mu$), but not necessarily the average baryon density ($n\equiv A/V$ where $V$ is the unit cell volume of the Coulomb lattice and $A$ is the mass number of the nucleus that is contained therein), are continuous functions. At $T=0$, the Gibbs free energy per baryon ($\mu$) and the energy of the system per baryon ($\varepsilon$) are related to the pressure and average baryon density as follows,
\be
\mu(A,Z;P,B)=\varepsilon(A,Z;P,B) + \frac{P}{n} \ .  \label{gibbs}
\ee
The composition ($A$, $Z$) of the outer crust for a fixed value of pressure and magnetic field ($B$) is determined by minimizing the Gibbs free energy per baryon $\mu(A,Z;P,B)$. Note that pressure, energy and density are related via $P=n^2 \frac{\partial \varepsilon}{\partial n}$ for a fixed number of baryons. Hence, the only unknown in the above equation is $\varepsilon(A,Z;P,B)$. The energy per baryon of the system, $\varepsilon(A,Z;P,B)$, is the sum of three independent contributions: 
\be \varepsilon(A,Z;P,B)=\varepsilon_n(A,Z)+\varepsilon_e(A,Z;P,B)+\varepsilon_l(A,Z;P,B), ~\label{eqn:Energy_ThreeTerms} \ee
namely the {\it nuclear}, {\it electronic} and {\it lattice} energy terms, respectively. 
In what follows, we will use natural units with $\hbar=c=1$, as is customary in the references that we have quoted.

In the absence of any magnetic field, the electronic contribution $\varepsilon_e(A,Z;P,B)$ would be that of a degenerate Fermi gas of relativistic electrons: as is well known, the Coulomb interaction among electrons becomes negligible at the densities we are considering. However, when a uniform magnetic field $B$ is present -- directed along the $z$-axis in our case --, while the energy keeps a continuous dependence on the component $p_z$ or the electron momentum component along the magnetic field, the energy levels are quantized and their values read \cite{Rabi1928,1930ZPhy...64..629L}
\be
E^2(\nu,p_z)=p_z^2+m_e^2 \left( 1 + 2 \nu B_\star\right) \ .
\label{eqn:FermiLevels}
\ee
Here, $m_e$ is the electron rest mass, $\nu$ is a non-negative integer quantum number and $B_\star\equiv B/B_c$, with $B_c\equiv\frac{m_e^2c^4}{e\hbar c}\approx 4.4\times 10^{13}$ G being the magnetic field at which the electron cyclotron energy becomes equal to the electron rest mass. 


For a fixed $\mu_e$ value, the maximum number of Landau levels populated by the electrons, named $\nu_\mm{max}$, can be evaluated by setting $E(\nu_\mm{max}, 0) = \mu_e$ in Eq.~(\ref{eqn:FermiLevels}), where $\mu_e$ is the electron chemical potential:  

\be \nu_\mm{max} = \mathrm{int}\left( \frac{\mu_e^2-m_e^2}{2 m_e^2 B_\star} \right) \ , \ee 
where ``int'' means the integer part.
In other words, the energy due to the electron coupling with the magnetic field cannot exceed $\mu_e$. Given that, the electron density $n_e$ is written as~\cite{Chamel2012}

\be
n_e=\frac{B_{\star}m_e^3}{2 \pi^2} \sum_{\nu=0}^{\nu_{\mathrm{max}}}{g_\nu \sqrt{ \left(\frac{\mu_e}{m_e}\right)^2  -1-2\nu B_\star}} \ ,
\ee
where $g_\nu=1$ for $\nu=0$ and $g_\nu=2$ for $\nu\neq 0$. 

Regarding the lattice energy $\varepsilon_l(A,Z;P,B)$, it is known to be independent of the magnetic field \cite{vleck1932}. Thus, results obtained for $B=0$ can be safely used here \cite{baym1971}. 
In addition, nuclear finite size effects, and zero-point energy contributions to the lattice energy have been found to be negligible \cite{Pearson2011}; therefore, we have not considered these effects.
The most energetically favorable configuration for the outer crust is the crystallization of the nuclei into a body-centered cubic lattice~\cite{FetterWalecka:2003}. The lattice energy per baryon can be written as
\be
\varepsilon_l = - C_l x^2 y^2 \overline p \ ,
\ee
where $C_l$ is a dimensionless constant, $C_l = 3.40665 \times 10^{-3}$ for a body-centered-cubic lattice in our units, $x\equiv A^{1/3}$ is proportional to the size of the nuclei at the vertices of the lattice, $y\equiv Z/A$ is their proton fraction and $\overline p\equiv (3\pi^2 n)^{1/3}$  is an auxiliary variable, that would represent the Fermi momentum of a uniform gas of free spin $1/2$ fermions with density $n$.

In our approach, $\varepsilon_n(A,Z)$ does not contribute to the pressure and it is not affected by the magnetic field. This latter statement is certainly valid for magnetic field strengths lower than $10^{17}$ G~\cite{Pena:2011, Wei:2024huk}. As shown in \cite{Basilico:2015ypa, Wei:2024huk}, for some typical nuclei appearing in the outer crust, the presence of $B\approx 10^{17-18}$ G produces a small nuclear extra-binding, favoring the stability of the nuclei present in the lattice. Due to this, the effect of $B$ on the nuclear binding is not expected to impact our qualitative conclusions concerning the synthesis of SHEs and the underlying mechanism favoring their appearance. 

For the sake of completeness, in Appendix \ref{app}, we give the set of equations needed to obtain the optimal composition of the outer crust, explaining briefly the method for the solution. More details can be also found in \cite{Basilico:2015ypa, Chamel2012}.

In Sec.~\ref{sec:res}, we will adopt two state-of-the-art nuclear models to account for the nuclear energy per baryon ($\varepsilon_n$). In particular, we will consider two Energy Density Functionals (EDFs), one relativistic and one non-relativistic.

\subsection{The case of large fields and large densities}

In the current section, we resort to a simplified nuclear mass model to shed some light, in a clear and transparent way, on the mechanism triggering the appearance of SHEs. For this illustrative purpose, we have adopted the well-known Liquid Drop Model (LDM), according to which the nuclear energy per baryon can be written as follows 
\begin{eqnarray}
  \varepsilon_n(x,y) &=& m_py + m_n(1 - y) + \varepsilon_v + \varepsilon_s + \varepsilon_c + \varepsilon_{\rm asym} \nonumber \\
  &=&m_py + m_n(1 - y) - a_v + \frac{a_s}{x} + a_c x^2 y^2 \nonumber\\
  &+& a_a(1 - 2y)^2 \ .
  ~\label{eqn:en_LDM}
\end{eqnarray}
The coefficients $a_v = \SI{15.71511}{MeV}$, $a_s = \SI{17.53638}{MeV}$, $a_c = \SI{0.71363}{MeV}$, $a_a = \SI{23.37837}{MeV}$ are associated with the volume term $\varepsilon_v$, the surface term $\varepsilon_s$, the Coulomb interaction term $\varepsilon_c$ and the asymmetry term $\varepsilon_{\rm asym}$, respectively. The numerical values are taken from Ref.~\cite{Roca:2008}.

As already mentioned, the presence of SHEs has been associated with extremely large magnetic fields,  $B\gtrsim 10^{18}$ G,  where $\nu_{\mathrm{max}}=0$ (strongly quantizing magnetic fields, cf.  \cite{Chamel2012}), and with average baryon densities in the innermost part of the outer crust of the order of $10^{-2}$ fm$^{-3}$ (cf. Fig.~3 in Ref.~\cite{sekizawa2023}). In these conditions, $\mu_e \gg m_e$. Hence, the electron chemical potential can be approximated by
\be
\mu_e \approx \frac{2\pi^2 n_e}{m_e^2B_\star}=\frac{2\pi^2 yn}{m_e^2B_\star}\equiv \frac{2}{3}y\frac{\overline p^3}{m_e^2B_\star} \ ,  
\ee
where we have checked that, in this range, the latter expression is accurate within 2\% with respect to the exact value.

Recalling that the electron contribution to the Gibbs energy per baryon is equal to $y\mu_e$ and that $P_l/n = \varepsilon_l/3$ \cite{Chamel2012}, one can write the Gibbs energy per baryon as a function of $\overline p$, 
\begin{eqnarray}
  \mu(x,y;\overline p,B_\star) &\approx& m_py + m_n(1 - y) \nonumber \\
  &-& a_v + \frac{a_s}{x} + a_c x^2 y^2 + a_a(1 - 2y)^2 \nonumber \\
  &+& \frac{2}{3}y^2\frac{\overline p^3}{m_e^2B_\star}-\frac{4}{3}C_l x^2y^2\overline p \ .
\end{eqnarray}
The latter expression allows one to easily realize that for a {\it critic} $\overline p^c \approx 157$ MeV (i.e., for density $n^c \approx 1.7\times 10^{-2}$ fm$^{-3}$), the nuclear Coulomb and lattice terms cancel each other $\left(a_C-\frac{4}{3}C_l \overline p^c\right)x^2y^2\approx 0$. In such a situation,  
\begin{eqnarray}
  \mu(x,y;\overline p^c,B_\star) &\approx& m_py + m_n(1 - y) \nonumber \\
  &-& a_v + \frac{a_s}{x} + a_a(1 - 2y)^2 \nonumber \\
  &+& \frac{2}{3}y^2\frac{\left(\overline p^c\right)^3}{m_e^2B_\star} \label{eqn:mu_pF} \ .
\end{eqnarray}
As a matter of fact, we emphasize that the role of the lattice term outside these extreme conditions (i.e. $n^{\rm drip} \lesssim 10^{-4}$ fm$^{-3}$) is known to be quite limited $\frac{4C_l\overline p^{\rm drip}}{3a_C}\lesssim 0.2$.

In order to find the optimal composition according to Eq.~(\ref{eqn:mu_pF}), one must write $\mu(x,y;\overline p^c,B_\star)$ as a function of the pressure. To this aim, we first relate the pressure with $\overline p$ using Eq.~(29) of Ref.~\cite{Chamel2012}, 
\be
P_c\approx \frac{y^2}{9\pi^2}\frac{\left(\overline p^c\right)^6}{m_e^2 B_\star} \ ,
\ee
which is valid for strongly quantizing magnetic fields, and large densities (or
$\mu_e \gg m_e$). In this regime, we have validated the accuracy of this expression which neglects the lattice contribution to the pressure. By writing 
Eq.~(\ref{eqn:mu_pF}) as a function of the pressure, we obtain
\begin{eqnarray}
  \mu(x,y;P_c,B) &\approx& m_py + m_n(1 - y) \nonumber \\
  &-& a_v + \frac{a_s}{x} + a_a(1 - 2y)^2 \nonumber \\
  &+& 2\pi\frac{y}{m_e}\left(\frac{P_c}{B_\star}\right)^{1/2} \ .  \label{eqn:mu_P}
\end{eqnarray}
From this equation, we first note that all the terms depend only either on the proton fraction $y=Z/A$ or on the mass number $x=A^{1/3}$. This implies that the optimal value of $y$ will be independent of $x$ and vice versa. Specifically, the optimal value of $x$ can be obtained by minimizing $\mu(x,y;P_c,B)$ with respect to $x$:
\be
0 = \frac{\partial \mu}{\partial x} =-\frac{a_s}{x^2} \ .   
\ee
This implies that $x=A^{1/3}\rightarrow \infty$, clearly showing
the underlying mechanism triggering the appearance of SHEs. 
We stress again that this result can be only realized thanks to the presence of the lattice contribution. When considering microscopic calculations for $\varepsilon_n$, as in Sec.~\ref{sec:res}, this scenario may not be precisely reached, since the cancellation of nuclear Coulomb and lattice terms may not be perfectly realized. 

Regarding the proton fraction $y$, since the asymmetry term $a_a$, that favors $y \rightarrow 1/2$, is not opposed by the nuclear Coulomb term,  that favors $y\rightarrow0$, the process of neutron enrichment in this region of the crust is expected to slow down. The limit imposed for the proton fraction in this simple model would be 
\begin{eqnarray}\label{yldm}
  0 = \frac{\partial \mu}{\partial y} &=&m_p-m_n - 4(1-2y)a_a+\frac{2\pi}{m_e}\left(\frac{P_c}{B_\star}\right)^{1/2},  \nonumber \\
  y&\approx&\frac{1}{2} - \frac{\pi}{4 m_e a_a}\left(\frac{P_c}{B_\star}\right)^{1/2}, \nonumber\\
  y&\approx&\frac{1}{2}\frac{1}{1+\frac{1}{12}\frac{\left(\overline p^c\right)^3}{m_e^2a_a B_\star}}\approx 0.3. 
  \label{eq:limit_y}
\end{eqnarray}
Here, the proton-neutron mass difference has been neglected. As we show below, this simple model leads to reliable qualitative values for $x$ and $y$, in its applicability regime.

\begin{figure}[t!]
    \centering
    \includegraphics[width=0.45\textwidth]{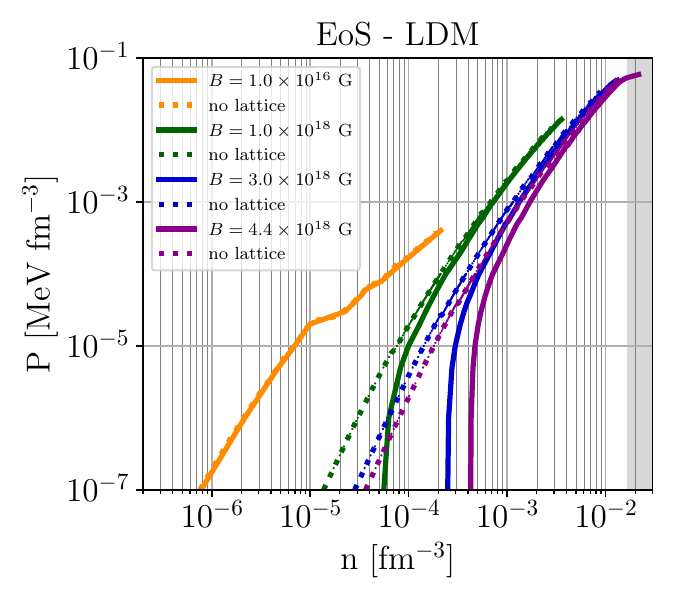}
    \caption{Equation  of state ($P$ versus $n$) predicted by the LDM, for different external magnetic field values ($B = 1.0 \times 10^{16} \, \mm G$, $B = 1.0 \times 10^{18} \, \mm G$, $B = 3.0 \times 10^{18} \, \mm G$, $B = 4.4 \times 10^{18} \, \mm G$), including (solid lines) and neglecting (dashed lines) the contribution of the lattice Gibbs energy. The curves are plotted up to the neutron-drip transition point.
    }
    \label{fig:EOS_LDM}
\end{figure}

\begin{figure}[t!]
    \centering
    \includegraphics[width=0.45\textwidth]{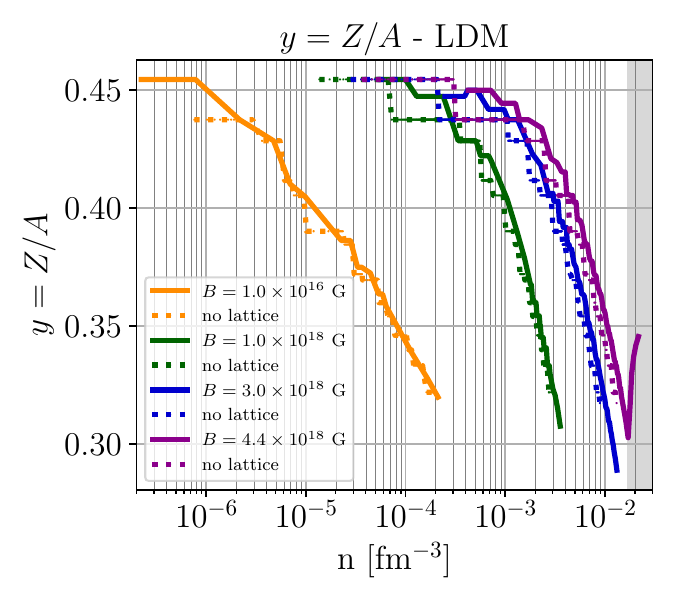}
    \caption{The $y=Z/A$ ratio versus $n$, 
    as predicted by the LDM model, for different external magnetic field values ($B = 1.0 \times 10^{16} \, \mm G$, $B = 1.0 \times 10^{18} \, \mm G$, $B = 3.0 \times 10^{18} \, \mm G$, $B = 4.4 \times 10^{18} \, \mm G$).
    The results either include
    (solid lines) or neglect the contribution of the lattice Gibbs energy (dashed lines). Curves are plotted up to the neutron-drip transition point.}
    \label{fig:ZoverA_LDM}
\end{figure}

\begin{figure*}
    \centering
    \includegraphics[width=\textwidth]{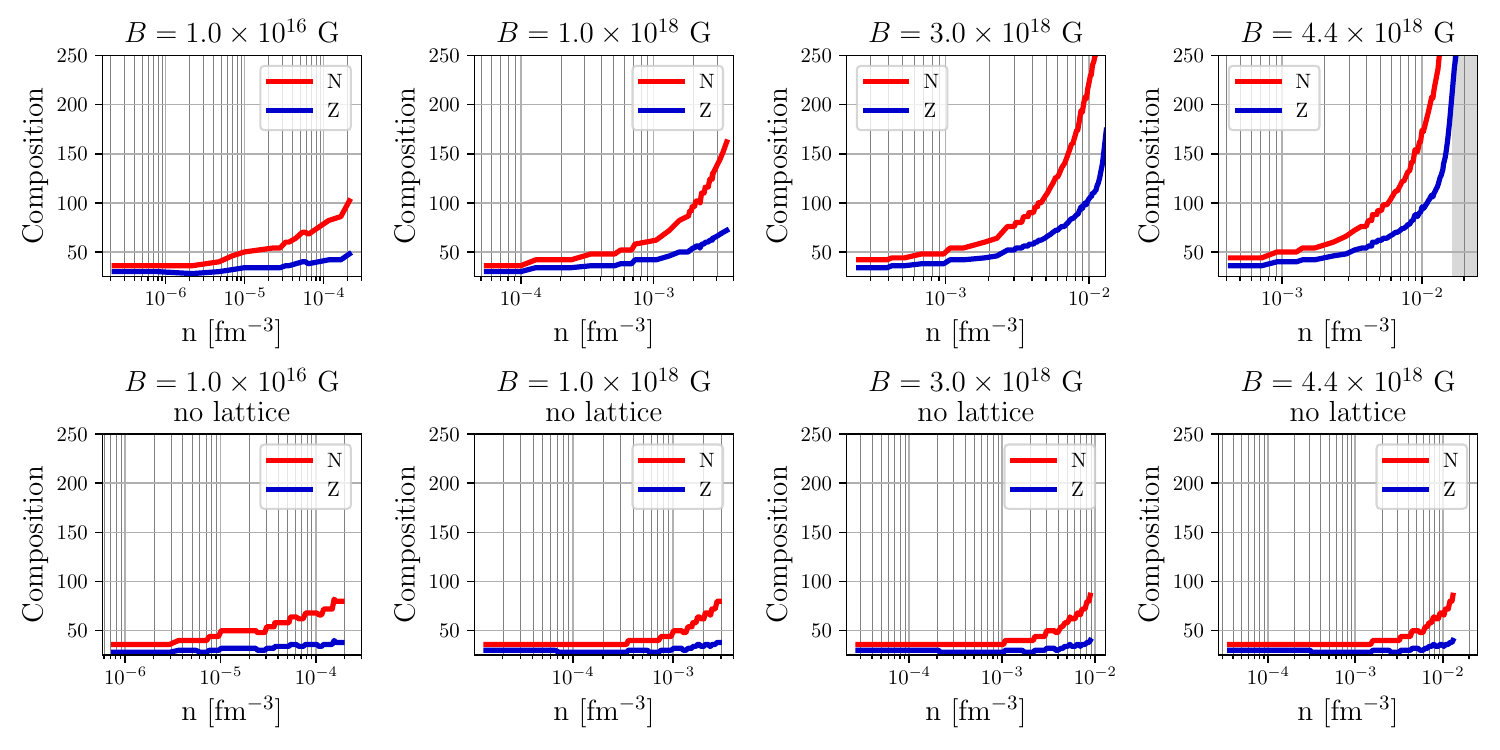}
    \caption{Composition, i.e. $Z(n)$ and $N(n)$ trends (blue solid line and red solid line respectively) of the outer crust of a magnetar  obtained by employing the LDM for the nuclear binding energies, for the following values of magnetic field: $B=\SI{1.0e16}{G}$, $B=\SI{1.0e18}{G}$, $B=\SI{3.0e18}{G}$, $B=\SI{4.4e18}{G}$.
    Curves are plotted up to the neutron-drip transition point.}
    \label{fig:Composition_LDM}
\end{figure*}

In Fig.~\ref{fig:EOS_LDM} we show the equation of state -- the pressure $P$ as a function of the number density $n$ --  predicted by the LDM  for $B = \SI{1.0e16}{G}$, $B = \SI{1.0e18}{G}$, $B = \SI{3.0e18}{G}$, $B=\SI{4.4e18}{G}$, without any approximation  
in the calculation of
$\mu(x,y;P,B)$ (solid lines). Results neglecting the Coulomb lattice are also displayed (dashed lines). It is evident that a large $B\gtrsim 10^{18}$ G induces outer crusts with larger densities and pressures before the neutron drip line is reached, as already 
pointed out e.g. in Refs.~\cite{Basilico:2015ypa,Chamel2012,ChamelN.2015RoLq,2023PhRvD.107d3022P,Chamel2020b,sekizawa2023}. In the outermost layer, for the highest magnetic fields where only the $\nu=0$ electronic level is filled, the density remains almost unchanged for a large range of pressure values, but this does not occur when neglecting the lattice contribution to the pressure. That is, the lattice makes the crust practically incompressible in the outermost layers (cf. Sec.~IV.a of Ref.~\cite{Chamel2012}). We note, however, that the present model is not strictly valid for describing the surface of the star because of the non-uniformity of the electron gas there~\cite{Lai2001}. In addition, effects due to finite temperature, not considered here, can considerably change the equation of state~\cite{Thorolfsson1998}.  In brief, for this low density and pressure range, our estimate of the equation of state should be considered as only qualitative.
Finally, we point out that in Fig.~\ref{fig:EOS_LDM}, as well as in the following Figs.~\ref{fig:ZoverA_LDM} and~\ref{fig:Composition_LDM}, we highlight the critical density $n^c$ by shadowing the region above it.

In Fig.~\ref{fig:ZoverA_LDM}, we show the ratio $y=Z/A$ 
in order to confirm by the numerical results our prediction in Eq.~(\ref{eq:limit_y}).
For the sake of completeness, we also show in Fig.~\ref{fig:Composition_LDM} our results for the composition of the outer crust, i.e. $Z(n)$ and $N(n)$, either including (upper panels) or neglecting (lower panels) the lattice contribution. The results are obtained by considering $B=\SI{1.0e16}{G}$, $B=\SI{1.0e18}{G}$, $B=\SI{3.0e18}{G}$, $B=\SI{4.4e18}{G}$, from left to right. It is clear, by comparing the upper with the lower panels, that the presence of the lattice contribution is responsible for the strong increase of both $Z$ and $A$ in the innermost layers of the outer crust, leaving approximately the $Z/A$ ratio unchanged (shown in Fig.~\ref{fig:ZoverA_LDM}). This is a more complete numerical validation of the analytic model that has been given above. 

It is important to note that, moving from the lowest magnetic field to the most intense one, the neutron drip density increases from $\approx 10^{-4}$ fm$^{-3}$ to $\approx 10^{-2}$ fm$^{-3}$, in agreement with the literature (see the above references). Nonetheless, this latter value is still far from the density range at which the pasta structures are expected to appear, that is around $n \approx 0.06-0.08 \, \mm{fm^{-3}}$. For very large magnetic fields, although smaller than the ones used here, the density at which the pasta phase is expected to appear increases with the magnetic field ~\cite{Scurto:2022vqm,Wang:2022sxx}. Accordingly, we can reasonably conclude that there is no apparent reason to expect pasta configurations in the enlarged outer crust that we find in our work.

A complete study of the transitions from isolated nuclei to the inner crust, and from the inner crust to possible pasta phases, is beyond our scope; it may be interesting in future to investigate how the large magnetic fields impact them, in a systematic way.


\section{Results}
\label{sec:res}

The possible appearance of SHEs was not considered in our previous work \cite{Basilico:2015ypa} due to the limits imposed on the proton and neutron numbers in the employed mass tables. Moreover, in that
work we did not consider magnetic fields whose values extend up to $B\sim 3-5 \times 10^{18}$ G. 

In this section, we present our results obtained from the numerical minimization of the Gibbs energy per particle, $\mu(A,Z; P, B)$ [cf. Eq.~(\ref{gibbs})], focusing on the composition based on two state-of-the-art nuclear mass models covering now a larger range of $Z$ values and, consistently, of $N$ values. Specifically, we have employed the relativistic  
DDPC1~\cite{DDPC1} and the non-relativistic 
UNEDF1~\cite{UNEDF1} energy density functionals.  
These mass tables can be found in \cite{NMT} including the evolution of the binding energy for nuclei up to $Z\sim 120-140$, and covering isotopes from the proton to the neutron drip lines (in the vacuum). In this section, we will pay special attention to the effects produced in the innermost layers of the outer crust, if extremely strong magnetic fields are present. 
We shall assess how these outcomes are qualitatively independent of the choice of the model to describe nuclear masses and consistent with the expectations of the previous Section.

\begin{figure*}
    \centering
    \includegraphics[width=\textwidth]{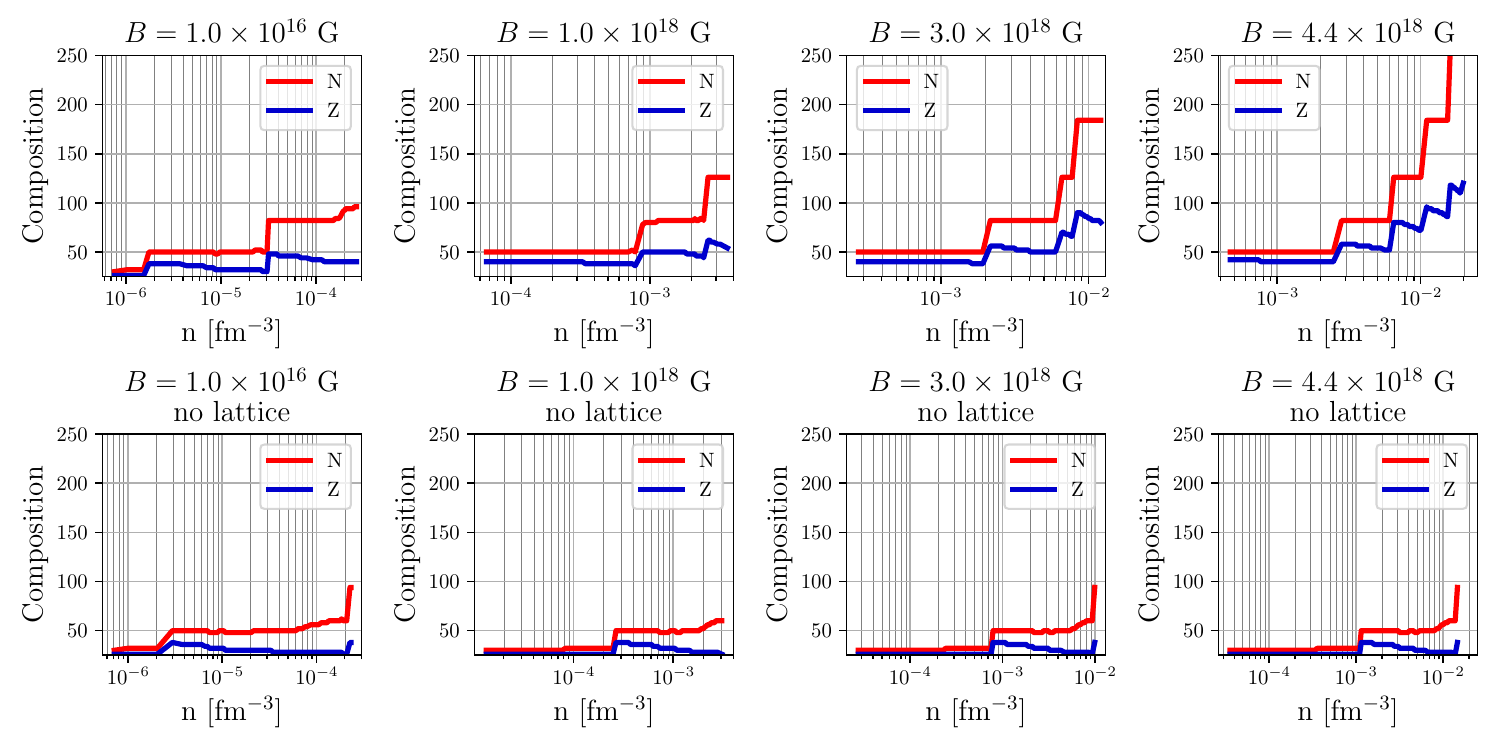}
    \caption{Compositions, i.e. $Z(n)$ and $N(n)$ trends (blue solid line and red solid line respectively) of the outer crust of a magnetar obtained by employing UNEDF1 model for the nuclei binding energies, for the following values of magnetic field: $B=\SI{1.0e16}{G}$, $B=\SI{1.0e18}{G}$, $B=\SI{3.0e18}{G}$, $B=\SI{4.4e18}{G}$.}
    \label{fig:Composition_UNEDF1}
\end{figure*}

\begin{figure*}
    \centering
    \includegraphics[width=\textwidth]{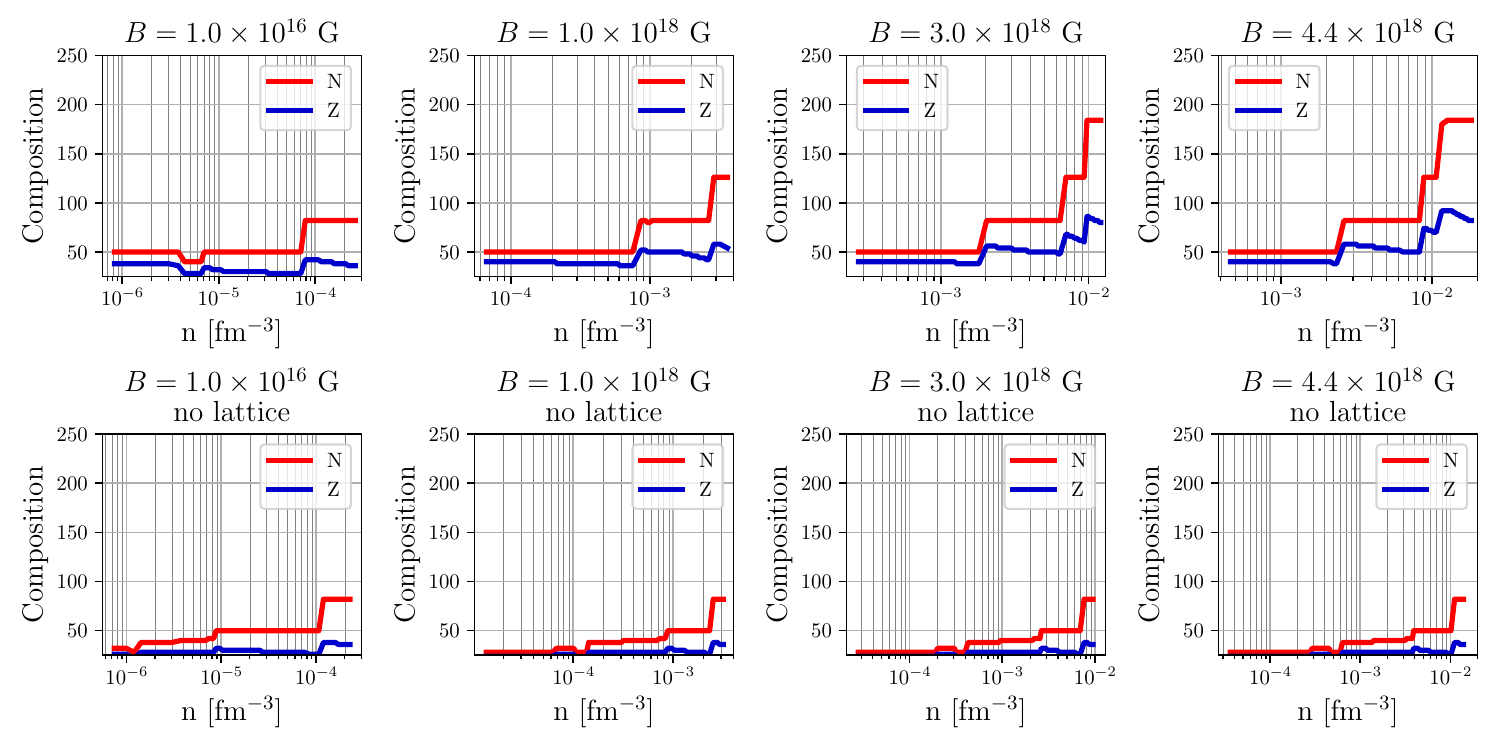}
    \caption{Compositions, i.e. $Z(n)$ and $N(n)$ trends (blue solid line and red solid line respectively) of the outer crust of a magnetar obtained by employing the DDPC1 model for the nuclei binding energies, for the following values of magnetic field: $B=\SI{1.0e16}{G}$, $B=\SI{1.0e18}{G}$, $B=\SI{3.0e18}{G}$, $B=\SI{4.4e18}{G}$.}
    \label{fig:Composition_DDPC1}
\end{figure*}

The nuclear composition of the outer crust, i.e. the functions $Z(n)$ and $N(n)$, are displayed in Figs.~\ref{fig:Composition_UNEDF1} and~\ref{fig:Composition_DDPC1} for the four different magnetic field values previously adopted, covering the same range explored in Ref.~\cite{sekizawa2023}. In these figures, we show results corresponding to the UNEDF0 (non-relativistic) and DDPC1 (relativistic) energy density functionals. As in the case of the LDM, the results obtained by neglecting the lattice contribution are shown in the lower panels.

For the lowest magnetic field, $B\sim \SI{1.0e16}{G}$, there are no significant differences with respect to previous works and the results obtained by neglecting the lattice contribution. Below $B\approx \SI{e18}{G}$, we reach maximum values of $Z\approx 60$ and $N\approx 130$, with slight variations according to the chosen nuclear mass model (in agreement with our previous results \cite{Basilico:2015ypa}). The presence of super-heavy and very neutron-rich nuclei emerges from $B\sim \SI{e18}{G}$ and it is evident for the highest analyzed magnetic field $B= \SI{4.4e18}{G}$, which induces the appearance of much larger densities as previously discussed. Specifically, SHEs appear for densities around $n\sim \SI{e-2}{fm^{-3}}$ as expected from our previous discussion based on the LDM. The aforementioned results are similar for the considered mass models, suggesting that they are robust against the choice of the model, and that the physics mechanism enabling the appearance of SHEs is qualitatively well understood on the basis of the simplified model previously introduced. Indeed, as we have already stressed in the case of the LDM, the calculations neglecting the lattice contribution (lower panels of Figs.~\ref{fig:Composition_UNEDF1} and~\ref{fig:Composition_DDPC1}) do not predict SHEs to appear in the outer crust.  

\begin{figure}[t!]
    \centering
    \includegraphics[width=0.45\textwidth]{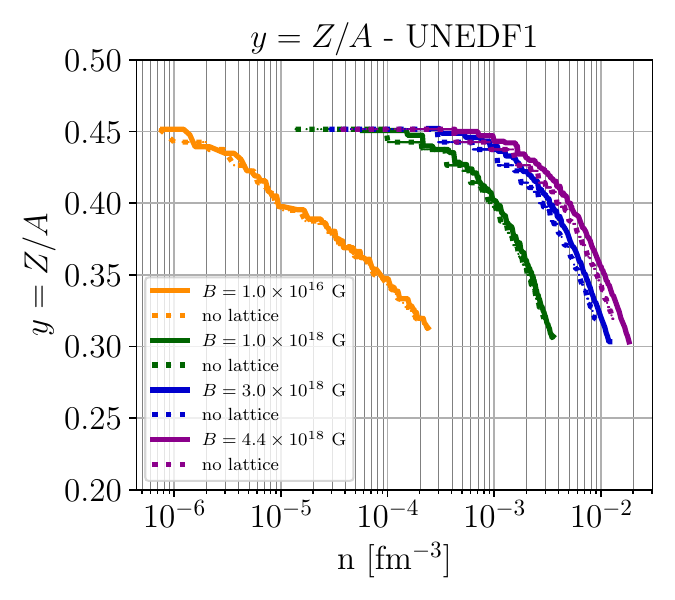}
    \caption{$y=Z/A$ ratio versus $n$ predicted by the UNEDF1 model, for different external magnetic field values ($B = 1.0 \times 10^{16} \, \mm G$, $B = 1.0 \times 10^{18} \, \mm G$, $B = 3.0 \times 10^{18} \, \mm G$, $B = 4.4 \times 10^{18} \, \mm G$), including (solid lines) and neglecting the contribution of the lattice Gibbs energy (dashed lines). Curves are plotted up to the neutron-drip transition point.}
    \label{fig:ZoverA_UNEDF1}
\end{figure}

\begin{figure}[t!]
    \centering
    \includegraphics[width=0.45\textwidth]{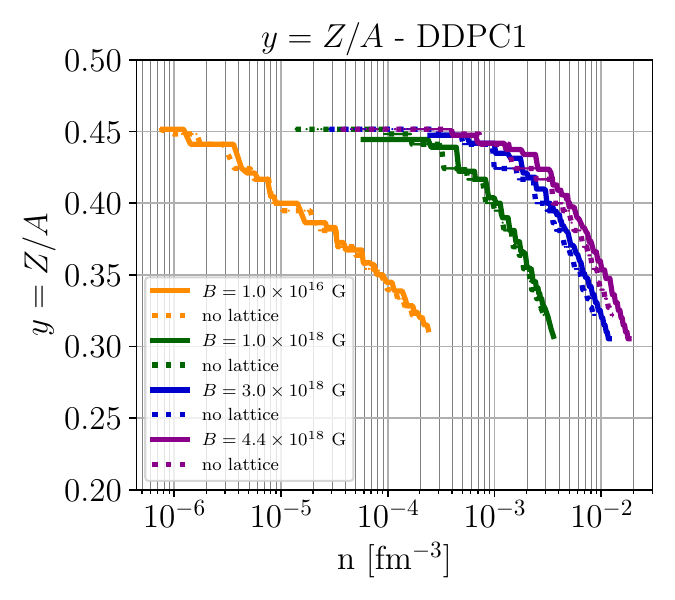}
    \caption{$y=Z/A$ ratio versus $n$ predicted by the DDPC1 model, for different external magnetic field values ($B = 1.0 \times 10^{16} \, \mm G$, $B = 1.0 \times 10^{18} \, \mm G$, $B = 3.0 \times 10^{18} \, \mm G$, $B = 4.4 \times 10^{18} \, \mm G$), including (solid lines) and neglecting the contribution of the lattice Gibbs energy (dashed lines). Curves are plotted up to the neutron-drip transition point.}
    \label{fig:ZoverA_DDPC1}
\end{figure}

\begin{figure}[t!]
    \centering
    \includegraphics[width=0.45\textwidth]{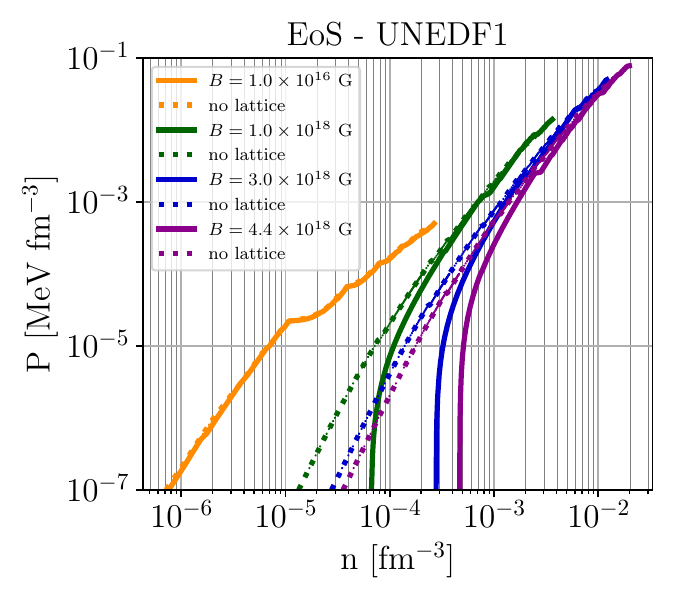}
    \caption{Equation of state ($P$ versus $n$) predicted by the UNEDF1 model, for different external magnetic field values ($B = 1.0 \times 10^{16} \, \mm G$, $B = 1.0 \times 10^{18} \, \mm G$, $B = 3.0 \times 10^{18} \, \mm G$, $B = 4.4 \times 10^{18} \, \mm G$), including (solid lines) and neglecting the contribution of the lattice Gibbs energy (dashed lines). Curves are plotted up to the neutron-drip transition point.}
    \label{fig:EOS_UNEDF1}
\end{figure}

\begin{figure}[t!]
    \centering
    \includegraphics[width=0.45\textwidth]{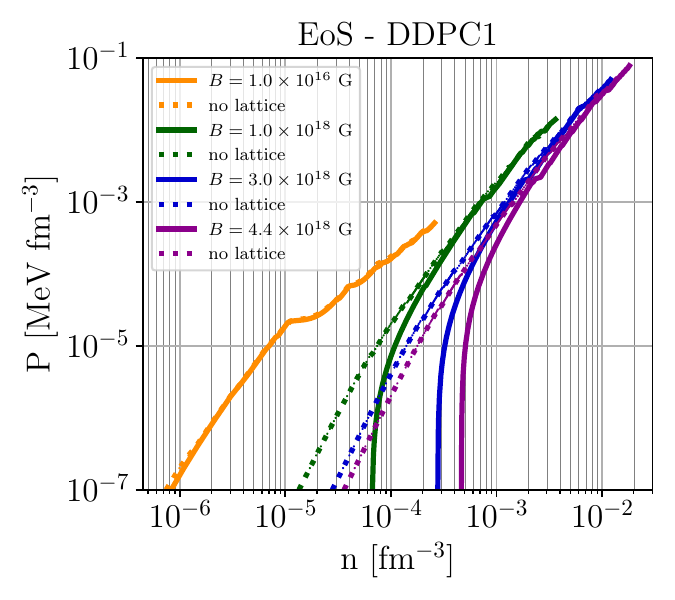}
    \caption{Equation  of state ($P$ versus $n$) predicted by the DDPC1 model, for different external magnetic field values ($B = 1.0 \times 10^{16} \, \mm G$, $B = 1.0 \times 10^{18} \, \mm G$, $B = 3.0 \times 10^{18} \, \mm G$, $B = 4.4 \times 10^{18} \, \mm G$), including (solid lines) and neglecting the contribution of the lattice Gibbs energy (dashed lines). Curves are plotted up to the neutron-drip transition point.}
    \label{fig:EOS_DDPC1}
\end{figure}

Another interesting feature of our results -- also predicted by the LDM model -- is that SHEs are not particularly unbalanced in the $y=Z/A$ ratio, as shown in Fig.~\ref{fig:ZoverA_UNEDF1} and Fig.~\ref{fig:ZoverA_DDPC1} for the UNEDF1 and DDPC1 models, respectively. The limiting value of $y$ is very similar to the outcome of Eq.~(\ref{eq:limit_y}). For the sake of completeness, we display in Figs.~\ref{fig:EOS_UNEDF1} and~\ref{fig:EOS_DDPC1} the equations of state associated with the UNEDF1 and DDPC1 mass models for $B = 1.0 \times 10^{16} \, \mm G$, $B = 1.0 \times 10^{18} \, \mm G$,  $B = 3.0 \times 10^{18} \, \mm G$ and $B = 4.4 \times 10^{18} \, \mm G$ (solid lines). Results neglecting the Coulomb lattice are also displayed (dashed lines). 
There are small differences between the results of UNEDF1 and DDPC1, although they cannot be visible on the overall scale.
Similar results have previously been found \cite{Chamel2012,ChamelN.2015RoLq,Basilico:2015ypa,2023PhRvD.107d3022P,Chamel2020b,Wei:2024huk}. We have checked that at least two more mass models that have been extended up to $Z\sim 120-140$ and cover isotopes from the proton to the neutron drip lines (in the vacuum), namely HFB-32~\cite{HFB} and UNEDF0~\cite{UNEDF0}, provide similar results for the crust compositions.

\section{Conclusion}
\label{sec:con}

In the present work, we have analyzed again, and in more detail, the nuclear composition of the outer crust of neutron stars that are characterized by extremely strong magnetic fields (magnetars). This topic has been already the subject of different works and yet, for the first time, the work of Ref.~\cite{sekizawa2023} has found, unexpectedly, that when those magnetic fields are larger than $\approx 10^{18}$ G, SHEs can be found in the innermost layers. In our current paper, we have proposed a clear explanation for this mechanism and discussed further important aspects.

First, we have confirmed the existence  of SHEs  by using  two 
state-of-the-art nuclear mass models. 
A physical interpretation for the synthesis of such elements is given in a transparent way, based on the LDM. Realistic calculations qualitatively follow the trends expected by our simplified analysis using the LDM, and confirm the insight that is gained thanks to this robust, albeit simple, 
model. 

In essence, if ever present, magnetic fields above $10^{18}$ G quantize the energies of the electrons, and allow them to reach larger momenta, compared to the case of the free electron gas. This enables the presence of larger densities, of about $10^{-2}$ fm$^{-3}$, within the outer crust.  We have clearly shown that these densities lead to an unusual situation where the contribution of the Coulomb lattice to the average chemical potential tends to quench, or even to cancel, the effect of the nuclear Coulomb term. This situation gives more freedom to the asymmetry term, which opposes more efficiently the increase with density of the electronic contribution to the average chemical potential, by slowing down the neutron enrichment of the outer crust. The role played by the lattice term to the Gibbs energy per baryon, for the largest densities and magnetic fields analyzed, turns out to be decisive. Eventually, in this situation, the most convenient way to optimize the chemical potential is to reduce the nuclear surface energy by allowing nuclei to be heavier and heavier.

Although this mechanism cannot be trivially replicated in terrestrial laboratories, it sheds some interesting light on the physics of SHEs and their synthesis. The very concepts of nuclear stability, drip lines and/or highest possible atomic number, are strongly altered in the medium that may exist in magnetars, provided magnetic fields of the order of $\approx 10^{18}$ G can be reached.

\begin{acknowledgments}
We would like to thank Prof. Sekizawa for bringing to our notice the possible appearance of superheavy elements in the outer crust of extremely magnetized neutron stars. We would like to thank Prof. Haskell for the useful discussions about the magnetars models and the understanding of the associated phenomenology.
XRM acknowledges support by MICI-U/AEI/10.13039/501100011033 and by FEDER UE
through grants PID2023-147112NB-C22; and through
the “Unit of Excellence Maria de Maeztu 2025-2028” award to the Institute of Cosmos Sciences, grant CEX2024-001451-M. Additional support is provided by the Generalitat de Catalunya (AGAUR) through grant 2021SGR01095.
\end{acknowledgments}

\appendix
\section{Algorithm to find the composition of the crust}
\label{app}

In a nutshell, to determine the outer crust composition, that is to seek for the optimal nucleus, we 
solve consistently the following set of equations, that determine $\mu_e$, $\nu_{\mathrm{max}}$, $n_e$, $p_z(\nu)$~\cite{Basilico:2015ypa}:
\begin{equation}
    \begin{dcases*}
        \mu_e^2=m_e^2\left( 1+2\nu_{\mathrm{max}} B_{\star} \right), \\
        p_z (\nu)^2+m_e^2 \left( 1+2\nu B_{\star} \right) = \mu_e^2, \quad  0\leq \nu \leq \nu_{\mathrm{max}}, \\
       n_e=\frac{B_{\star}m_e^3}{2 \pi^2} \sum_{\nu=0}^{\nu_{\mathrm{max}}}{g_\nu \sqrt{ \left(\frac{\mu_e}{m_e}\right)^2  -1-2\nu B_\star}}   , \\
        P=P_e+P_l(A,Z). \\
        \end{dcases*} ~\label{eqn:foureq}
\end{equation}
Since the four coupled equations above cannot be solved analytically, we have adopted a numerical procedure to solve Eq.~(\ref{eqn:foureq}) for a given magnetic field $B$ and for a fixed pressure $P$, which can be summarized as follows. First of all, we enter a tentative chemical potential $\mu_e$. From this, we extract $\nu_\mm{max}$, $n_e$, and the $(\nu + 1)$ values of $p_z(\nu)$ from the first three equations. By inserting these quantities and the input values of $Z$ and $A$, $P_e+P_l$ is determined. The best $\mu_e$ value is the one making the right-hand term of the fourth equation equal to the input pressure. As a consequence, the Gibbs energy (Eq.~\ref{gibbs}) can be calculated. This procedure is repeated for each $Z$ and $A$ values of interest, and the best pair of values $Z$, $A$ are those ones minimizing the Gibbs energy reported in Eq.~(\ref{gibbs}). The neutron-drip transition is determined by the condition $g(P_\mathrm{drip}) = m_n$: it corresponds to the condition for which it is energetically favorable for the system to start dripping neutrons from the nucleus and form a neutron gas.

\bibliographystyle{apsrev4-2}
\bibliography{references.bib}

\end{document}